

\input phyzzx

\line{
\hfill ULB-TH 7/93}
\line{\hfill gr-qc/9404025}
\line{
\hfill April 1993}
\bigskip
\centerline  {\bf ENTROPY GENERATION IN QUANTUM GRAVITY}
\centerline {\bf AND BLACK HOLE REMNANTS\foot{presented by F.Englert}}
\bigskip
\centerline{ A. CASHER and F. ENGLERT\foot{E-mail: casher at taunivm;
fenglert at ulb.ac.be}}
\medskip
\centerline{\it Service de Physique Th\'eorique}
\centerline{\it Universit\'e Libre de Bruxelles, Campus Plaine, C.P.225 }
\centerline{\it Boulevard du Triomphe, B-1050 Bruxelles, Belgium}
\centerline{\it and}
\centerline{\it School of Physics and Astronomy}
\centerline{\it Raymond and Beverly Sackler Faculty of Exact Sciences}
\centerline{\it Tel-Aviv University, Ramat-Aviv, 69978 Tel-Aviv, Israel}
\bigskip
\bigskip
\noindent
{\bf Abstract}

The area entropy $A/4$
 and the related Hawking temperature in the presence of
event    horizons are
 rederived, for de Sitter and black hole topologies, as a
consequence of a tunneling
 of the wave functional associated to the classical
coupled matter and gravitational fields.  The extension of the
wave functional outside the
 barrier provides a reservoir of quantum states which
allows for an additive constant
 to $A/4$. While, in a semi-classical analysis,
 this gives no new information
  in the de Sitter case, it yields   an infinite
constant in the black hole case. Evaporating black holes would then leave
residual ``planckons" - Planckian
 remnants with infinite degeneracy. Generic
planckons can  neither
 decay into, nor be directly
  formed from, ordinary matter in a
finite time.  Such   opening
    at the Planck scale of an infinite Hilbert space is
expected to provide the ultraviolet
 cutoff required to render the theory
  finite in the
sector of large scale physics.
 \vfil \eject \noindent
{\bf 1. Introduction}
\medskip
Tunneling in quantum gravity can generate entropy$^{[1],[2]}$. To
understand how
such an apparent violation of unitarity may arise, let us first
consider a classical spacetime background geometry with compact Cauchy
hypersurfaces. If quantum fluctuations
 of the background are taken into
account,
quantum gravity leaves no ``external"  time parameter to describe the
evolution
of matter configurations in this background. Indeed, the solutions of the
Wheeler-De Witt equation$^{[3]}$  $${\cal H} \vert \Psi \rangle = 0 \eqno
(1) $$
where $\cal H$ is the Hamiltonian density  of the interacting
gravity-matter
system can contain no reference to such time when there are  no
contribution to
the energy from surface terms at spatial
 infinity$^{[4]}$. This is due to the
vanishing
of the time displacement generator and even though the theory can be
unambiguously formulated only at the semi-classical level, such a
consequence of
reparametrization invariance should have
 a more general range of validity.

To parametrize evolution, one then needs
 a ``clock" which would define time
through correlations$^{[5]}$, namely correlations
 of matter configurations with
ordered sequences of spatial geometries. If quantum
 fluctuations of the metric
field can be neglected, the field components $g_{ij}$
 at every point of space can
always be parametrized by a classical time
parameter, in accordance with the classical equations of motion. This
classical
time, which is in fact a function of the $g_{ij}$, can  be used to
describe the evolution of matter and constitutes thus such a dynamical
``clock"
correlating matter to the gravitational field$^{[6]}$.
 This description is
available in the Hamilton-Jacoby limit of (1) where the
 classical background
  evolving in time is represented by a coherent superposition of
``forward" waves  formed from eigenstates of (1). When
 quantum metric field
fluctuations are taken into account, ``backward" waves, which   can  be
interpreted as flowing backwards in time, are unavoidably generated from
(1)
and the operational significance of the metric clock gets
 lost outside the
domain of validity of the Hamilton-Jacoby limit.
Nevertheless, in domains of metric field configurations where
 both forward
and
backward waves are present but where quantum fluctuations are
 sufficiently
small, interferences with such ``time reversed" semi-classical
 solutions
will in
general be negligible\foot{For a recent discussion of related
 problems see
reference $[7]$.}. Projecting then out the backward waves restores the
operational significance of the metric clock but the evolution marked by
the
correlation time is no more unitary: information has been lost in
projecting
these backward waves stemming from regions where quantum
 fluctuations of the
clock are significant. This is only an apparent violation of unitarity
which
would be disposed of if the full  content of the theory would be
kept, perhaps eventually by reinterpreting backward waves in terms of
the creation
of ``universe" quanta through a further quantization  of
 the wavefunction
(1).

This apparent violation of unitarity is particularly marked if the
gravitational
clock experiences the
 strong quantum fluctuations arising from a tunneling
process. This can be illustrated from the simple analogy, represented in
Fig.1,
offered by a nonrelativistic closed system of total fixed energy  where
a
particle in one space
 dimension $x$ moving in a potential $U(x)$ plays the role
of a clock for surrounding
 matter and tunnels through a large potential barrier.
Outside the barrier, the
clock is well approximated
  by semi-classical waves, but if on the left of
the
turning point $a$ one would
 take only forward waves, one would inevitably have
on
the right of the other turning
 point $b$ both forward and backward waves with
large
amplitudes compared with the original ones. The ratio between the squares
of the
forward amplitudes on the right and on the left of the barrier for a
component
of the clock wave with given
 clock energy $E_c$ is the inverse transmission
coefficient $N_0(E_c)$ through the barrier and provides a measure of the
apparent violation of unitarity. In the
Hamilton-Jacoby limit of
 quantum gravity, the characterisation of tunneling
amplitudes by inverse
 transmission coefficients $N_0$ will appear as the natural
one to compute the entropy
 transferable reversibly between the metric clock and
matter. More precisely, we
 shall see that, in this limit, spherically symmetric
spacetimes  bounded by event horizons are in
general connected by tunneling
 to another manifold and that  the entropy gained
by tunneling from the latter to
 the former is, for large barriers, $\log N_0$.
Explicit evaluation  of  this tunneling
 entropy yields $\ln N_0 = A/4$ where $A$
is the area of the event horizon. In this
 way the horizon thermodynamics of
Gibbons and Hawking$^{[8]}$ is recovered. But the present approach has
potentially   additional information.

The tunneling entropy $\ln N_0$ is in last analysis an effect
of quantum fluctuations in quantum gravity.
 Therefore, despite the fact
that
no violation of
 unitarity would  appear in a complete description including
backward waves, this entropy should be expressible in terms of density of
states of matter and gravity. Tunneling offers an interesting perspective
in this
direction because it enlarges the semi-classical wave function of
spacetime to
include in its description
 the other side of the barrier. This can yield a
reservoir of quantum states
 which may provide, in addition to the $\exp (A/4)$
states building the entropy,
 residual states which would be expressed as an ``integration
constant" in the total entropy ${\cal S}$ of spacetime. Thus we shall
write $$ {\cal S} = A/4 + C \eqno(2) $$
and try to get some information
 about the constant $C$ by analysing both sides
of the barrier.

The knowledge of $C$ is crucial, in particular for the
understanding    of the black hole
 behaviour at the final stage of evaporation.

A infinitely large value of $C$ would
 indeed indicate that the evaporation can
only radiate a finite
number of ``surface" states of order
 $\exp (A/4)$ out of an infinite set of
available internal states. This mismatch would entirely modify the black
hole
evaporation process at its last stage and bring the decay to a halt.
Indeed when the black hole   evaporates
 to the  Planck scale, it becomes, if $C$
is infinite,  a ``planckon"$^{[9]}$, that
 is a  remnant with infinite degeneracy. Causality and
unitarity prevent
 the decay and the
  production in a finite time of   planckons directly out of
ordinary matter for
 nearly all such states$^{[9]}$. Namely, if a  planckon state $\vert A_i
\rangle $ of finite size and mass $m$ decays (or is produced)  within
a finite time $\tau$
 in an approximately flat space-time background, the total
number of
possible final (or initial)
states is limited through causality
by the number ${\cal N}\left( \tau \right)  $
  of orthogonal
states
with total mass $m$ in a volume $\tau^3$. Assuming the number of
quantum fields which describe physics at scales large compared to the
Planck scale   to be finite, ${\cal N}\left( \tau \right) $
is a finite number.
Unitarity then implies that if the dimension $ \nu\left( m\right) $ of
the Hilbert space spanned by the degenerate states
                     $ \vert A_i \rangle$    becomes greater than
${\cal N}\left( \tau \right) $, a subspace of planckon states whose
dimension is $\nu \left( m\right) - {\cal N}\left(\tau \right) $
will be, for times smaller than $\tau$, orthogonal to the Hilbert
space of  states formed by these quantum fields. Thus, when
$\nu (m)
\to \infty$, generic
 planckons cannot decay (nor be formed) in a finite
time. Of course the above
 argument does not preclude the very formation of the
{\it finite} number of distinct
 planckons which can be generated in a finite
time as remnants of macroscopic
 decaying black holes. This time is however
unrelated to the time for their
 decay (creation)   directly at the Planck size
into (from) ordinary matter
 quanta; the latter time is generically  infinite.

 On the
other hand, a zero or finite
 value of $C$ would lead to the disappearance of the
hole at the end point of evaporation
 and hence probably imply a genuine violation
of unitarity within our universe\foot{For
 a comprehensive review on recent
attempts to solve the black hole unitarity puzzle,
 see reference ${[10]}$. See
also reference ${[11]}$.}.

The main content of our work is that, in an asymptotically
 flat background,
the constant $C$ for black holes, as deduced from a WKB
analysis of tunneling, is in fact  infinite. Hence the present
approach indicates that    the solution of the unitarity problem posed by
the black hole decay is
 provided by planckon remnants. This conclusion is however
contingent upon the limitation of the semi-classical approach to quantum
gravity used here and remains therefore a tentative one.

We shall first review the computation of tunneling amplitudes in quantum
gravity through static
 barriers$^{[2]}$ and compute the  tunneling entropy  for
the   case of de Sitter spacetime topologies. However the estimation of
$C$ appears in this case
 intractable within the semi-classical approximation. We
then shall analyse in similar
 terms the black hole geometries$^{[12]}$. The
above-mentioned results will be derived and discussed. \bigskip

\noindent
{\bf 2. Tunneling amplitudes in quantum gravity}
\medskip
Our basic action in four dimensional
Minkowski space-time will be $$S=S_{grav}+S_{matter}\eqno(3)$$
where $ S_{grav}$ has the conventional form ($G=1$) :
$$ S_{grav}=-{1\over 16\pi} \int \sqrt{-g}R\, d^4x \eqno (4)$$
and $S_{matter}$ contains sufficiently many free parameters  to allow for
the
stress tensors considered
 below. A possible cosmological constant term can be
included in the matter action.

 Consider (Fig.2) in general two spacelike
hypersurfaces $\Sigma_1$ and $\Sigma_2$ which are turning points in
superspace
(or turning  hypersurfaces) along which   solutions  of the Minkowskian
classical equations of motion for gravity and matter meet a classical
solution
of their Euclidean extension. $\Sigma_1$ and $\Sigma_2$ are thus the
boundaries
of a   region $\cal E$ of  Euclidean space-time defined by the Euclidean
solution.
If $\cal E$ can be continuously shrunk to zero one can span $\cal E$
by a continuous set of hypersurfaces $\tau_e=$ constant such that $\tau
_e\equiv
\tau_{e,1}$ on
  $\Sigma_1$ and $\tau_e \equiv \tau_{e,2}$ on  $\Sigma_2$. These
$\tau_e=$
constant surfaces define a Euclidean coordinate system which we shall call
synchronous; the
Euclidean metric in $\cal E$ can be written in the form $$ds^2 =
N^2(\tau_e,x_k)\,
d \tau^2_e+g_{ij}
(\tau,x_k)\,dx^i\, dx^j \eqno(5)$$ where $N(\tau_e,x_k)$ is a
lapse
function. The Euclidean  action $S_e$ over $\cal E$, from $\Sigma_1$ to
$\Sigma_2$, is obtained by analytic continuation from the Minkowskian
action
(3) and can be written as
 $$\eqalign{S_e(\Sigma_2,\Sigma_1)   =\int_{\cal E}\Pi^{ij}
g_{ij}^\prime\,d^4x +\int_{\cal E} \Pi^a  \Phi_a^\prime\, d^4x
&-  \int_{\cal E}    (g_{ij} \Pi^{ij})^\prime \,d^4x\cr  &-{1\over
8\pi}\int_{\cal E}
 \partial_k[(\partial_j N)g^{kj}\sqrt {g^{(3)}}]\,d^4x. \cr}
\eqno (6) $$
 Here $\Pi^{ij}$ and $\Pi^a$ are the Euclidean momenta conjugate to the
gravitational
 fields $g_{ij}$ and to the matter fields  $\phi_a$; $g^{(3)}$
is
the three dimensional determinant and the $\prime$ symbol indicates a
derivative with
 respect to $\tau_e$. In the gauges (5),  $\Pi^{ij}$ is expressed
as
  $$ \Pi^{ij}= {\sqrt{ g^{(3)}}\over
32\pi N}[g^{im}g^{jn}-g^{ij}g^{mn}] g_{mn}^\prime,
\eqno(7)$$

On the turning hypersurfaces $\Sigma_1$ and $\Sigma_2$, all field
momenta
 $(\Pi^{ij},\Pi^a)$ are zero in the synchronous system and the third term
in (6)
vanishes.  The last term in (6) also vanishes if the
hypersurfaces $\Sigma_1$ and $\Sigma_2$ are compact   but may receive
contributions from infinity otherwise. In this case, we shall assume that
turning hypersurfaces merge at infinity  sufficiently fast so that
the Euclidean action $S_e(\Sigma_2,\Sigma_1)$ does not get contributions
 from the last term in (6).  The classical Minkowskian solution in the
space-time ${\cal M}_1$ containing $\Sigma_1$ can be represented quantum
mechanically by a ``forward wave" solution $\Psi(g_{ij},\phi_a)$ of the
Wheeler-De Witt equation (1) in the Hamilton-Jacoby limit. At $\Sigma_1$,
this
wave function
 enters, in the WKB limit, the Euclidean region ${\cal E}$ and
leaves it at
 $\Sigma_2$ to penetrate a new Minkowskian space-time ${\cal
M}_2$.
The tunneling of $\Psi(g_{ij},\phi_a)$ through  ${\cal E}$ engenders in
addition
to the  ``forward wave" solution a time reversed ``backward wave". The
inverse
  transmission coefficient $N_0$ through the barrier  measures the
ratio
of the norms
 of the forward waves at  $\Sigma_2$ and  $\Sigma_1$. For large
$N_0$ one may write in the synchronous system
$$ N_0 = \exp {\left[-2(\int_{\cal
E}\Pi^{ij}    g_{ij}^\prime\,d^4x +\int_{\cal E} \Pi^a
 \phi_a^\prime \, d^4x)\right]}. \eqno (8) $$
As all surface terms in (6) vanish in this
 system, (8) can be rewritten in the coordinate invariant form
$$
N_0 = \exp {[2S_e(\Sigma_1,\Sigma_2)]}. \eqno (9) $$

Let us examine
 the case where the Euclidean manifold  ${\cal E}$ is static
in the sense that it
admits a
 Killing symmetry.    We can take advantage of the covariance of
the   action $S_e$ and express it in terms of a new ``static" coordinate
system, possibly
 singular,  with momenta everywhere vanishing in  ${\cal E}$. In
this way, momenta
 in (8) get squeezed into the last surface term of (6) and
one gets  $$ N_0=   \Delta t_e {1\over
4\pi}\int_{\cal E} \partial_k[(\partial_j N)g^{kj}\sqrt {g^{(3)}}]\,d^3x
\eqno(10) $$
 where $\Delta t_e$
  is the Euclidean time needed to span ${\cal E}$ in the
static system. The
 tunneling amplitude will then be computable from this surface
term only, even when
 the static parametrisation is singular. \bigskip
\noindent
{\bf 3. The tunneling entropy in de Sitter spacetime topologies}
\medskip
Let us first illustrate
 the equivalence implied by (9) of the expressions (8)
and (10) for the de Sitter spacetime which is the classical solution of
pure gravity in the presence
 of a cosmological constant $\Lambda$.  In the
present formalism, $\Lambda$
 should be viewed as the Lagrangian density of the
matter action in (2); it plays
 the role of a matter distributed with rest energy
density $\sigma = \Lambda$ and
 obeying the equation of state $\sigma =- p$ where
$p$ is a (negative) pressure.
The full Minkowskian solution is the 4-hyperboloid
(Fig.3) which can be parametrized by the minisuperspace metric
$$ ds^2 = d\tau^2
-a^2 d\sigma^2;\quad a=r_h \cosh {\tau \over r_h} \eqno(11)$$
 where $r_h =(3/8\pi
\Lambda)$.
 The hypersurface $\tau=0$ is a turning hypersurface connecting the
hyperboloid
 to the  Euclidean
  solution consisting of the 4-sphere which can be described in
a synchronous system
 by replacing in (11) $ \tau$ by $-i\tau_e$. This yields the
Euclidean scale factor
$$ a_e= r_h \cos {\tau_e \over r_h}. \eqno(12)$$
 The
half-sphere delimited
 by $-\pi r_h/2 \leq \tau_e \leq 0 $ has another turning
point at, say, the south
 pole $\tau_e = -\pi r_h/2$ where $ a_e=0$\foot {In
fact, any point on the
 half-sphere can be taken as a turning hypersurface.}: in
the above synchronous
 system the space integral of the momenta in a $\tau_e =$
constant hypersurface
 vanishes in the vicinity of this point and so does the
third term in (6).
  The half-sphere considered constitute the domain ${\cal
E}$ through which a
 ``wormhole" at, say, $a_e=0$ is connected by tunneling to the
de Sitter spacetime. The inverse transmission coefficient $N_0$ can be
straightforwardly
 computed from (8)  using (7) and one gets $$ N_0 = \exp \left[
{3\pi \over 2} \int_{-\pi r_h/2}^{+\pi r_h/2} a_e \left({da_e \over
d\tau_e}\right)^2\,
 d\tau_e \right]= \exp (\pi r_h^2) = \exp (A/4) \eqno (13) $$
where $A$ is the area of the event horizon.

The significance of
 this result is best appreciated when the 4-sphere is
described in static coordinates:
$$ ds^2= (1-r^2/r_h^2)\,
 dt_e^2 + (1-r^2/r_h^2)^{-1}\, dr^2 + r^2\, d\Omega^2.
\eqno(14) $$
In this static frame, all momenta vanish everywhere on the sphere and the
tunneling is expressible by the surface term (10) only where the radial
integration is carried
 from $r=0$ to $r=r_h$. The Euclidean time is periodic with
period ${\cal T}^{-1} =
 2\pi r_h$. Using (10) with $\Delta t_e = (1/2){\cal
T}^{-1}= \pi r_h$, one recovers the result (13).

It is now easy
to verify that the equality
 between the inverse transmission coefficient and
$\exp (A/4)$ is maintained when the de Sitter spacetime is perturbed by
spherically symmetric static
 matter distributions$^{[2]}$. This establishes
the   validity of (13) for these generalized de Sitter spacetimes.

Let us tentatively take boundary
  conditions in field space by assigning pure
forward  waves at the wormhole turning point. The  probability of finding
an expanding generalized de Sitter
 spacetime for a corresponding wormhole state
is then $N_0$, since in the classical limit interferences between
spaces evolving forward or backward
 in time must be negligible. Assuming that all
wormhole states are equally probable, we get from (13) that the relative
probability of finding two matter
 configurations in the generalized de Sitter
spacetimes is
$$ {N_0^{(1)} \over N_0^{(2)}}=
 \exp \left[ { A^{(1)}\over 4}- {A^{(2)}\over 4}
\right] .    \eqno(15)$$
Integrating
 the constraint equation ${\cal H}=0$ over a static domain of the
Minkowskian spacetime one gets
$$ {1\over 16\pi}
 \int \sqrt{-g}R\, d^3x + H_{matter}  + { {\cal T} A\over 4}
= 0 \eqno(16)$$
where $H_{matter}$
 is the total matter energy. The variation of (16) yields
$$ - { \delta A \over 4 }={\cal T}^{-1}\delta_\lambda
H_{matter} \eqno(17) $$
 where $\lambda$
  labels the {\it explicit} dependence of  $H_{matter}$ on all
other (non gravitational) ``external" parameters. Equation (17) is the
differential Killing identity of reference [13].

It now follows from (15) and (17)  that matter configurations
with neighbouring
 energies in a static patch of a generalized de Sitter
spacetime would be Boltzmann
distributed at the
 global temperature ${\cal T}$  provided our ignorance about
wormhole states allows to take them to be  equally probable. Thus the
temperature of the
 static patch is  $\cal T$ and therefore (17) also implies
that  $ A/4$
is (up to an integration
 constant $C_{de Sitter}$) the   entropy of spacetime
and that the latter is in thermal
equilibrium with the surrounding matter. As the
entropy must be an intrinsic property of spacetime, not
only is equilibrium a consequence of the chosen boundary conditions in
field
space but the converse is
 also true: the temperature obtained directly from
(17)
with the spacetime
 entropy identified as $A/4 + C_{de Sitter} $ must agree  at
equilibrium with the thermal
distribution generated from the field boundary conditions. This justifies
a posteriori the above choice of boundary conditions\foot{up to changes
which
would not alter the probability ratios in the large $N_0$ limit.}.

The tunneling approach to the horizon entropy and temperature$^{[1],[2]}$
used  here   differs from the analysis based on the
Euclidean
periodicity of Green's
 functions$^{[8]}$ in two respects. On the one hand,
the
present approach yields the thermal spectrum, and then the entropy, from
the
{\it backreaction\/}  of the thermal matter on the gravitational field,
in
contradistinction to the Green's function approach. On the other hand
however,
the thermal matter considered here is taken in the classical limit while
the
Green's function
 method describes genuine quantum radiation. Both methods fall
short of a fully consistent quantum treatment of the backreaction.
But as stated in
 the introduction the present approach may uncover from the
hidden side of the barrier a density of state building the full entropy.
Unfortunately, for
 the de Sitter spacetimes considered above, the hidden side is
a wormhole whose description cannot be achieved in our semi-classical
approach. Hence, for
 de Sitter spacetimes, we do not gain at this stage any
information on the
 integration constant $C_{de Sitter}$  which measures the
density of states left when the full spacetime reduces to the (planckian)
wormhole. As we shall
 now see the situation appears quite  different in the case
of black hole geometries. \bigskip
\noindent
{\bf 4. The tunneling
 entropy of black holes}
\medskip

A Schwarzschild static
 patch of an eternal black hole of mass $m_0$
is described by the metric
$$ds^2= (1-{2m_0\over r})\, dt^2 - (1-{2m_0\over r})^{-1}\, dr^2 - r^2\,
d\Omega^2. \eqno (18) $$
Surrounding the black hole by static matter generalizes
 (18) to
$$ ds^2 = g_{00}(r)\,dt^2
- g_{11}(r)\, dr^2 - r^2\, d\Omega^2 \eqno (19) $$
where in absence of outer horizon one has
$$r\to
\infty: g_{00}(r) = g_{11}^{-1}(r) \to 1-{2MJ\over r}. \eqno(20) $$
Here $M$ =$M(\infty)$ is the total mass and
$$\eqalign {M(r) &= m_0 + \int_{2m_0}^r 4\pi z^2 \sigma (z)J\,dz \cr
g_{11}^{-1}(r) &= 1-{2M(r)J\over r}\cr
g_{00}(r) &= \left( 1-{2M(r)J\over r}\right) \exp \left[- \int_r^\infty
(\sigma + p_1) 8 \pi z g_{11}(z) \,dz \right] \cr}. \eqno(21) $$

The metric
 (19) can be extended to the four quadrants of a Kruskal space and we
choose identical
 matter distributions in the two Schwartzschild patches to keep
a twofold symmetry around the Kruskal time axis. The Kruskal diagram is
depicted in Fig.4
 where we have also indicated its Euclidean extension $T_e=
iT$ resulting from the analytic continuation of the static metric (19) to
the periodic time $t_e =it$. The Euclidean periodicity is
$${\cal T} = {1\over
4\pi}[g_{00}(2m_0)
\,g_{11}(2m_0)]^{-1/2} {dg_{00}(r)\over dr}\vert_{r=2m_0}
\eqno(22)$$
 or from (21)
 $$ {\cal T}= {1\over 8\pi m_0}
\exp{\left[-\int_{2m_0}^{\infty}{ (\sigma + p_1) 4\pi z  g_{11}\,
dz}\right]}. \eqno(23) $$
The Euclidean extension
of the black hole surrounded by static matter is represented in Fig.5.

In contradistinction
 to the de Sitter case, there is clearly no WKB tunneling
from a wormhole to a
 black hole because of the mismatch in topologies. One is
therefore lead to investigate possible tunnelings between two black holes
geometries $(B.H.)_1$
 and $(B.H.)_2$ constituted respectively by black holes of
mass $m_0$ and $m$ ($m>m_0$) surrounded by matter. The
Euclidean sections of  $(B.H.)_1$  and $(B.H.)_2$, depicted in Fig.6, are
engendered
 by a rotation of half a Euclidean period of the hypersurfaces $a_1$
and  $a_2$ labeled by $T=0$ in their Kruskal diagrams. These are
turning hypersurfaces
 along which Minkowskian and Euclidean black holes meet. We
now search for two   black holes such that $a_1$ and $a_2$ are also the
boundaries of an Euclidean
 solution ${\cal E}$ of the Euclidean equations of
motion through which tunneling
 can take place from one Minkowskian black hole
to the other. A necessary condition
 for this to happen is that the total mass $M$
of the two black hole-matter systems
 and their Euclidean period ${\cal T}^{-1}$
be the same, so that the turning hypersurfaces $a_1$ and $a_2$ of the
two geometries merge at spatial infinity.

Let us choose identical matter distributions
  outside a radius $r_c = 2m
+\eta$. $\eta $ is  positive and such that
 the mass of the matter between the
horizon and $r_c$ is $0$ for $(B.H.)_2$ and
 thus $m-m_0$ for $(B.H.)_1$. Keeping
$m$ and the matter distribution outside $r_c$
 fixed, we now decrease $m_0$
towards $0$. From (23),   in order to keep the Euclidean period ${\cal
T}^{-1}$ constant for $(B.H.)_1$, we have also
 to decrease $\eta$ towards $0$.
As $\eta \to 0$ the mass $m-m_0$ surrounding the infinitesimal mass $m_0$
in  $(B.H.)_1$ approaches its own Schwartzschild
 radius. It then follows from
(21) that $g_{00}(r)$ tends to zero for the whole
 interval $2m_0 < r < 2m$. In
other words any frequency stemming from the neighbourhood
 of the small mass hole
is infinitely
 redshifted by the matter in that interval, as encoded in the
damping exponentials
 in equations (21) and (23).

It is clear, from the
 static coordinate description (19) extended to Euclidean
times $t_e = it$, that
 the  Euclidean sections of $(B.H.)_1$ and $(B.H.)_2$
coincide for $r>2m + \eta$ but,
 while for $(B.H.)_2$ the Euclidean section
terminates at $r=2m$,    $(B.H.)_1$
 presents an extra ``needle" in the region
$2m_0< r<2m$ whose 4-volume is vanishingly
 small when  $\eta \to 0$. As we now
show, this is where tunneling between  $(B.H.)_1$ and $(B.H.)_2$ occurs.

To this effect, following the notations of
 section 2, we  identify at finite
$\eta$ the first turning hypersurface
 $\Sigma_1$ through which tunneling takes place  with $a_1$ and consider
instead of a second turning hypersurface  $\Sigma_2$ a hypersurface
$a_2^\prime$ which lies in the Euclidean section of both $(B.H.)_1$ and
$(B.H.)_2$; thus $r$ is greater than $2m +\eta$
 everywhere on $a_2^\prime$.
When $\eta \to 0$, we can choose  $a_2^\prime$  arbitrarily close to
$a_2$. One can then prove$^{[12]}$ that all gravitational
momenta vanish in this limit on  $a_2^\prime$ in
 a synchronous system.  We may
then identify $a_2^\prime$ with $\Sigma_2$. The region $\cal E$ is
thus contained in the needle $2m_0<r<2m+\eta$. Because of the
Kruskal twofold symmetry $a_1$ is mapped onto itself by a Euclidean time
rotation of half a period and thus $\cal E$ spans
 only half the needle 4-volume.
{}From (9), we learn that the inverse transmission
 coefficient $N_0$  is simply the
exponential of the total Euclidean action of the
 needle. Although the
limiting 4-volume of the needle vanishes, the action will turn out to be
finite. It is in fact computable as the difference between the
Euclidean actions of the two black holes  cut off
  at the arbitrary radius $r_c$
greater than $2m$ because the two geometries and the
 two actions coincide for
all $r >r_c$.

We thus write
$$N_0= \exp [S_e^{(B.H.)_2}-S_e^{(B.H.)_1}]. \eqno(24)$$
To evaluate these actions we take advantage of the
 covariance to express them
in terms of the static coordinate system (19) with $t=-it_e$. Using
equations (10) and (22) and the fact that the integrand
 in (10) is the same at
$r_c$ for $S_e^{(B.H.)_1}$ and for $S_e^{(B.H.)_2}$, we get
$$N_0= \exp {[4\pi m^2 - 4\pi m_0^2(\eta \to 0)]}  \eqno (25)$$
or, as $m_0$ vanishes in the limit,
$$N_0= \exp {A/4} \eqno (26) $$
where $A=16 \pi m^2$ is the area of the event horizon of the black hole.

We have thus learned that black holes are related by quantum tunneling
to another classical solution for gravity and matter, namely to
a ``germ black hole" of
 infinitesimal mass determining the spacetime topology
surrounded by a static distribution of matter characterized by a vanishing
$g_{00}(r)$. This domain of space-time is characterized  by a
limiting light-like Killing vector.  When the space-time geometry
presents a 4-domain endowed with such a  Killing vector, we shall
call the domain an ``achronon".
 All spherically symmetric achronon configurations
will exhibit  an infinite time dilation in the Schwarzschild time $t$,  or
equivalently massless modes emitted by the achronon  are infinitely
redschifted. Classically, the
 achronon has  the ``frozen" appearance  of a
collapse at infinite Schwarzschild
 time. The difference is that it is  also
frozen  in  space-time.

To see that achronons can indeed be
 constructed, at least in a
phenomenological fluid model, we shall
 build a shell model with the required
properties.

Let us consider a static spherically symmetric distribution of matter
surrounded by an extended shell comprised between two radii $r_a$ and
$r_b$. We define
$$ \hat\sigma
\equiv \int_{r_a}^{r_b}\sigma g_{11}^{1/2}dr, \quad \hat p_\theta \equiv
\int_{r_a}^{r_b}p_\theta g_{11}^{1/2}dr, \quad \hat p_1 \equiv
\int_{r_a}^{r_b}p_1 g_{11}^{1/2}dr \eqno(27)$$
where $p_{\theta}
 = -T_{\theta}^{\theta}$ and $p_{\phi}=-T_ {\phi}^{\phi}$.
Assuming $p_1=0$,
 one may perform the thin shell limit  $r_b \to r_a =R$ in
these integrals  using  (21) and the Bianchi identity
$$p_{\theta}= p_{\phi}= {1\over4}(\sigma + p_1) {8\pi r^2 p_1
+ 2M(r)/r \over 1-2M(r)/r}+{1\over2}rp^{\prime}_1+p_1. \eqno(28)$$
One then gets
$$ \eqalignno {4\pi R \hat\sigma &=(1-2m^-/R)^{1\over2} -
(1-2m/R)^{1\over2}
&(29)\cr 8\pi R \hat p_\theta & = {1-m/R \over (1-2m/R)^{1\over2}} -
{1-m^-/R
\over (1-2m^-/R)^{1\over2}};\quad \hat p_1 =0  &(30)\cr}  $$
where $m$ and $m^-$ are the values of $M(r)$ respectively at $r_b$ and
$r_a$
and $m_s = m - m^-$
 is thus the mass of the shell. Equations (29) and (30)
are
the standard result$^{[14]}$. As the radius $R$ approaches $2m$, these
solutions become physically meaningless when  $\hat p_\theta$  becomes
greater than $\hat\sigma$: this violates  the  ``dominant energy
condition"$^{[15]}$, implying the existence of observers for which the
momentum
flow of the classical matter becomes spacelike;   in fact, the shell is
mechanically unstable even before this condition is violated$^{[16]}$.

The divergence of $\hat p_\theta$ when
$R \to 2m$  appears in (30) because of the vanishing denominator in
(28).   Equation (30)
 depends however crucially on the radial pressure being
zero inside the shell. Relaxing this condition it is  possible to avoid
all singularities of
 the stress tensor as $R \to 2m$ by requiring $p_1$  inside
the shell to satisfy, before performing the thin shell limit,
$$ 4\pi r^2 p_1
+
{M(r)\over r}=0 .\eqno(31)$$
This solution is  unsatisfactory if the (extended) shell sits
in an arbitrary background because of the finite discontinuity of the
radial
pressure across the shell boundaries which would lead to singularities in
$p_\theta$. We may ensure continuity of the radial pressure by immersing
the
shell in suitable
 left and right backgrounds. To avoid reintroducing stress
divergences when $r^b$ approaches $2M(r^b)$ these should satisfy $(\sigma
+
p_1)=0$ at the shell
 boundaries. One can now perform the thin shell limit.
The finite discontinuity of $p_1(r)$ at $r =R$ leads now to
$$  \hat p_\theta = - \hat {\sigma \over 2},\quad
\hat p_1 =0\eqno(32) $$
instead of (30), while $ \hat \sigma$ is still given by
(29).  The dominant
 energy condition is now satisfied everywhere and provided
the
background is smooth enough in the neighbourhood of the
shell, no stress divergences will appear when it approaches the
Schwarzschild radius. Performing the explicit integration over the
shell in the exponential
 term in (21), one gets for $g_{00}(r)$in the region
$0 \leq r<2m$,
$$ g_{00}(r)=(1-{2M(r)\over r})\left[{R-2m \over R-2m^-}\right]
\exp{\left[-{\cal R}\int_r^{\infty}(\sigma + p_1) 8\pi z g_{11}\,
dz\right]}.\eqno(34) $$
Here the radius $R$ of the shell is taken at $R= 2m + \eta$ where
$\eta$ is a positive infinitesimal and the symbol $\cal R$ means that
the
integral is carried over the regular matter contribution only. Clearly, $
g_{00}(r) = O(\eta)$ for
 $0 \leq r<2m$, $t$ arbitrary and the above matter
distribution constitutes
 indeed an achronon. These phenomenological shell models
are designed   only to
 illustrate achronon properties but are hardly physically
relevant as such and we
 shall tentatively assume that realistic field theoretic
achronons do exist. This
 is of course a crucial issue which deserves further
study.

We now relate in general the tunneling as encoded by (26) to
the black hole entropy. This can be done following the analysis of the de
Sitter case. Assume all
 states formed by achronons of mass $m$ surrounded by
matter configurations of
 mass $M-m$, $M$ fixed, to be equally probable. This
amounts here to assume  the validity of the microcanonical ensemble as
achronons can be viewed
 just as lumps of ordinary matter taken out from the
surroundings. The relative probability of finding
two black
 hole geometries for a given total mass $M$ is then given by (15) with
$A^{(1)}$ and $A^{(2)}$ identified here with the black hole areas. The
differential Killing identity (17) follows as before from the integrated
constraint
 equation, the only difference being in general an additional term
$\delta M$ on the right hand side arising from a surface term at spatial
infinity. As $M$ is kept
 fixed, this term plays no role and (17) remains valid
as such. Therefore, in
 analogy with the de Sitter case, matter configurations
with neighbouring energies
 in a static Schwartszchild patch of an eternal black
hole surrounded by matter
 have a Boltzmann distribution at a global temperature
$\cal T$. The latter now coincides with the local temperature at spatial
infinity. $\delta A/4$ is
 the differential entropy of the hole and $A/4$ is the
amount of entropy transferable to matter reversibly. The total black hole
entropy is $$ {\cal S}  =
 A/4 + C_{B.H.} \eqno(34) $$ where $\exp C_{B.H.}$
measures the number of quantum
 states of a residual planckian  black hole. The
boundary condition in field space  at equilibrium are  such that the wave
functional of an   eternal black
 hole has a small amplitude   describing an
achronon configuration whose relative
 weight with respect to the classical black
hole configuration is of order $\exp -(A/8) $. \bigskip
\noindent
{\bf 5. From achronon to planckons}
\medskip
We now discuss the nature and the
significance of $C_{B.H.}$.
The entropy $A/4$ which can be exchanged reversibly from a black hole to
ordinary matter was rederived in the preceding section from the existence
of a
``potential barrier" between a black hole of mass $m$ and an achronon of
the
same mass.
 This was done in the context of   eternal black holes admitting
a
Kruskal twofold symmetry, so that there are in fact    two achronons
imbedded in two causally disconnected static spaces. Within each space
black
hole-achronon
 states are in thermal equilibrium with their surroundings. We
are
therefore led to picture in such a space a quantum black hole  state , in
the semi-classical
limit, as a quantum superposition of two coherent (normalized) states,
$\vert
B.H. \rangle$ and $\vert A \rangle$ representing respectively a classical
black
hole and a classical achronon. The
relative weight of the two states   is
approximately, up
to a phase, $\exp (-A/8)$ . It  follows from detailed balance at
equilibrium
between radiated matter and the black hole that the same superposition
should
hold for a the black hole who would only emit (and not receive) thermal
radiation at the equilibrium temperature. As a black hole formed from
collapse
would asymptotically
 emit such a thermal flux, we infer that the collapsing black
hole  approaches a state $\vert C
\rangle$   containing
 an achronon component with the same weight as in thermal
equilibrium. We thus
 write $$ \vert C \rangle \simeq \vert B.H. \rangle + \exp
(-A/8)\vert A \rangle. \eqno(35)$$

To a classical single black hole configuration one may associate many
distinct classical achronon configurations. In the shell model, for
instance,
there are infinitely many distinct classical matter configurations of the
same
total mass $m$. The argument is however much more general and   infinite
quantum degeneracy of
 the achronon is a direct consequence of the infinite
time
dilation. Indeed, the
 Hamiltonian $H_{matter}$ is of the form  $$ H_{matter}=
\int \sqrt{g_{00}}K(\phi_a , g_{ij}, \Pi_a,)\,d^3x \eqno(36)$$
and all its eigenvalues are squashed towards zero by the Schwarzschild
time
dilation factor $\sqrt{g_{00}}$, thus generating an infinite number of
orthogonal zero energy modes on top of the original achronon.

The infinity of zero energy modes around  any background implies an
infinite
degeneracy of achronons of given mass and thus an
infinity of distinct quantum black hole states of the same mass differing
by
the achronon component
 of their wave function. This infinite degeneracy of
the
quantum black hole  provides
 the reservoir from which are taken the finite
number of ``surface" quantum
 states $\exp A/4$ counted by the area entropy
$A/4$ transferable reversibly to outside matter.

Except for providing a rational
 for the large but finite testable entropy
of the
black hole, achronons do not modify the behaviour of large macroscopic
black
holes. However
 when their mass is reduced by evaporation and approaches the
Planck mass the barrier disappears and quantum superposition completely
mixes the
two components.
 Of course, this means that both the description in terms of
semiclassical
 configurations and of tunneling disappears. What remains however
as a consequence of unitarity, is the infinity of distinct orthogonal
quantum
states available
 which have no counterpart in the finite number of decayed
states. The quantum black hole  has become a planckon$^{[12]}$, that
is a Planckian mass
 object with infinite degeneracy. In terms of equation (34),
this means that in a
 asymptotically flat background, the integration constant of
the black hole
 entropy $C_{B.H.}$ is infinite. As discussed in the introduction,
this means that
 in such a background a generic planckon cannot decay nor be
formed in a finite time.

This conclusion
 however is contingent upon the validity at the qualitative
level of our semi-classical approach and upon the possibility of building
achronons from
 genuine field theory. The main question is whether or not the
{\it quantum}
 backreaction of the matter on the metric removes the infinite
degeneracy.
 Answering it requires further analysis.

Finally, it
 is of interest to note that the planckon solution to the unitarity
problem posed
 by the evaporating black hole would have, at a  fundamental level,
far reaching
 implications on the spectrum  of quantum gravity. The opening at the
Planck size
 of an infinite number of states, an unavoidable consequence of the
existence of
 planckons, may appear as a horrendous complication which could make
quantum gravity definitely unmanageable
but hopefully the converse may be true. Indeed planckons should make
quantum
gravity ultraviolet
 finite. The Hilbert space of physical states available
to
macroscopic observer must be orthogonal to the infinite set of states
describing
Planckian bound states. Their wave function at Planckian scales where
planckon
configurations are concentrated are therefore expected  to be vanishingly
small.
In this way, planckons would provide the required short distance cut-off
for a
consistent field theoretic description of quantum gravity within our
universe
while leaving the largest part of its information content hidden at the
Planck
scale.

An operational formulation of quantum gravity applicable within our
universe and
based on  conventional
 four dimensional gravity  might thus well be within
reach. Nevertheless, the sudden widening of the spectrum of physical
states
at the Planck scale and the relative scarcity of states
which describe large distance physics
 suggest that a fully consistent theory cannot be formulated in terms
of only long range quantum fields (including the metric),
and a larger scheme may be required
to cope with the infinite amount of information relegated to the
Planck scale.
\bigskip
\centerline{{\bf{Acknowledgements}}}

\noindent
We are very grateful to R. Balbinot, R. Brout, J. Katz, J. Orloff, R.
Parentani
and Ph. Spindel for
 most enjoyable, stimulating and clarifying discussions.

\vfil
\eject

\centerline{ {\bf References}}
\medskip
\item{[1]}
F. Englert,``From Quantum Correlations to Time and Entropy'' in ``The
Gardener
of Eden'' Physicalia Magazine (special issue in honour of R. Brout's
birthday),
(1990) Belgium, Ed. by P. Nicoletopoulos and J. Orloff.

\item{[2]}
A. Casher and F. Englert, Class. Quantum Grav. {\bf 9} (1992) 2231.

\item{[3]}
B. De Witt, Phys. Rev. {\bf 160} (1967) 113.

\item{[4]}
T. Regge and C. Teitelboim, Annals of Physics {\bf 88} (1974) 286.

\item{[5]}
D. Page and K.W. Wooters, Phys. Rev. {\bf D27} (1983) 2885. \hfill \break
Y. Aharonov
 and T. Kaufherr, Phys. Rev. {\bf D30} (1984) 368. \hfill \break
J.B. Hartle, Phys. Rev. {\bf D37} (1988) 2818; Phys. Rev. {\bf D38} (1988) 2985

\item{[6]}
T. Banks, Nucl. Phys. {\bf 249} (1985) 332, \hfill \break
R. Brout, Foundations of Physics {\bf 17} (1987) 603, \hfill \break
R. Brout, G. Horwitz and D. Weil, Phys. Lett. {\bf B192} (1987) 318.

\item{[7]}
W. Unruh and W.H. Zurek, Phys. Rev. {\bf D40} (1989) 1064, \hfill \break
J.J. Halliwell, Phys. Rev. {\bf D39} (1989) 2912.

\item{[8]}
G. Gibbons and S. Hawking, Phys. Rev. {\bf D15} (1977) 2738; 2752.

\item{[9]}
Y. Aharonov, A. Casher and S. Nussinov, Phys. Lett. {\bf B191} (1987) 51.

\item{[10]}
J.A. Harvey
 and A. Strominger, ``Quantum Aspects of Black Holes", Preprint
EFI-92-41; hep-th/9209055.

\item{[11]}
T. Banks, M.O'Loughlin and A .Strominger , ``Black Hole Remnants and the
Information Puzzle" Preprint RU-92-40; hep-th/9211030.

\item{[12]}
A. Casher and
 F. Englert, ``Black Hole Tunneling Entropy and the Spectrum of
Gravity" Preprint ULB-TH 8/92; TAUP 2017-92; gr-qc/9212010.

\item{[13]}
J.M. Bardeen,
 B. Carter and S.W. Hawking, Comm. Math. Phys. {\bf 31} (1973)
161.

\item{[14]}
J. Frauendiener, C. Hoenselaers and W. Conrad,  Class. Quantum Grav. {\bf
7}
(1990) 585.

\item{[15]}
S.W. Hawking and G.F.R. Ellis, ``The Large Scale of Space-Time", (1973),
(Cambridge University Press, Cambridge, England).

\item{[16]}
P.R. Brady, J Louko and E. Poisson, Phys. Rev. {\bf D44} (1991) 1891.
\vfil
\eject

\noindent
{\bf Figure Captions}
\medskip

\noindent
Figure 1. Tunneling of a nonrelativistic ``clock".

\noindent
The energy of the clock $E_c$ is represented by the dashed line. On the
left of
the turning point $a$ the clock is well represented by a forward wave
depicted here
 by a single arrow. On the right of the turning point $b$ the
amplification of the forward wave and the large concomitant backward wave
are
indicated.
\bigskip
\noindent
Figure 2. Tunneling in quantum gravity.

\noindent
The two solid curves represent turning hypersurfaces $\Sigma_1$ and
$\Sigma_2$ separating the dark gray Euclidean region $\cal E$ from two
Minkowskian spacetimes depicted in light gray.

\bigskip

\noindent
Figure 3. Tunneling in de Sitter topology

\noindent
The heavy solid
 line delineates a 4-hyperboloid and the thin one a wormhole.
The Euclidean
 domain $\cal E$ constituted by a half 4-sphere is delineated by a
dashed line. The dotted circle is the turning hypersurface $\tau = 0$.

\bigskip
\noindent
Figure 4. The Kruskal
 representation of a black hole, eventually surrounded
by static matter.

\noindent
The dashed straight
 line is the Euclidean axis $T_e$ and the  dashed circle is
the analytic continuation in Euclidean time of the solid hyperbolae
representing
trajectories $r=$constant in the static patches I and III. These are
separated
from the dynamical regions  II and IV by the horizons $r=2m_0$ where lay
the
past and future
 singularities $r=0$ depicted by the dashed hyperbolae. The
Schwarzschild time $t$ run on opposite directions on the two hyperbolae
$r=$constant
 and the Euclidean time $t_e$ spans the period ${\cal T^{-1}}$
on the
analytically continued circle.

\bigskip

\noindent
Figure 5. Euclidean black hole surrounded by static matter.

\noindent
Each point is a 2-sphere and the circles span the
Euclidean time
 $t_e$. The heavy solid line is the turning hypersurface described
in Kruskal time by $T=0$.
\bigskip
\noindent
Figure 6. Black hole  tunneling.

\noindent
The figure
 represents the Euclidean sections of the two black hole geometries
$(B.H.)_1$
 and $(B.H.)_2$. The $(B.H.)_1$  geometry is depicted by thick lines
and the $(B.H.)_2$
 geometry by thin lines in the region where it differs from the
first.

\noindent
The curve $a_1$ represents a turning hypersurface of $(B.H.)_1$, to be
identified with $\Sigma_1$.
 The curve $a_2$ represents a turning hypersurface of
 $(B.H.)_2$. The curve $a^{\prime}_2$ represents a hypersurface
 which
  lays in the intersection of  the Euclidean sections of $(B.H.)_1$ and
$(B.H.)_2$ and tends  to $\Sigma_2$ in the limit $m_0 \to 0$.

\vfill \eject


\end